\documentclass[%
 reprint,
 amsmath,amssymb,
 aps,
]{revtex4-2}

\usepackage{graphicx}
\usepackage{dcolumn}
\usepackage{bm}
\usepackage{hyperref}
\hypersetup{colorlinks=true, linkcolor=blue, citecolor=red, urlcolor=blue}
\usepackage{soul}
\usepackage[english]{babel}

\usepackage{amsmath,amsfonts,amssymb,amsthm,mathtools,array,color}

\usepackage{physics}

\begin{document}

\preprint{APS/123-QED}

\title{Dissipation-free modes in dissipative systems}

\author{Daan van Seters$^{1}$}
\author{Tim Ludwig$^{1}$}%
\author{H. Y. Yuan$^{1}$}%
\author{Rembert A. Duine$^{1,2}$}

\affiliation{%
$^{1}$Institute for Theoretical Physics, Utrecht University,\\
Princetonplein 5, 3584 CC Utrecht, The Netherlands
}

\affiliation{$^{2}$Department of Applied Physics, Eindhoven University of Technology,
P.O. Box 513, 5600 MB Eindhoven, The Netherlands}%

\date{\today}

\begin{abstract}
The coupling between a system and its environment (or bath) always leads to dissipation. We show, however, that a system composed of two subsystems can have a dissipation-free mode, if the bath is shared between the two subsystems. Reading in reverse, a shared bath does not contribute to the dissipation of all modes. As a key example, we consider a simple model for a two-sublattice antiferromagnet, where the environment is modeled by a bath that is shared between the two sublattice magnetizations. In our model, we find that the Néel order parameter is a dissipation-free mode. For antiferromagnets, our results offer an explanation for why the dissipation rate of the Néel vector is typically much lower than that of the average magnetization. In general, our results suggest a way to reduce dissipation (and decoherence) for some modes in composite systems, which could have experimental and technological applications.
\end{abstract}

\maketitle


\textit{Introduction.---}A key topic in modern science and technology is the manipulation of information. As information carriers, various options have been proposed, including electric charge \cite{BardeenPR1948}, electron spin \cite{AtsufumJMMM2020,YuanPR2022}, and nuclear spin \cite{BernhardRMP2013} in solid state systems. All information carriers have in common that an interaction with their environment is inevitable. The interaction with the environment leads to dissipation or, more specifically, to energy waste and information loss. For example, electron transport in transistors is accompanied by Joule heating, which significantly increases the transistor's energy consumption and reduces its thermal stability when scaled down to the nanoscale. As another example, consider a spin-qubit which encodes binary quantum information and is a building block for quantum computing and information processing \cite{NielsenBook2011}. However, the interaction of spin-qubit with surrounding electrons, lattices, and defects usually leads to relaxation and dephasing of the quantum states and, in turn, destroys the quantumness of the qubit \cite{BernhardRMP2013}. As these examples illustrate, reducing the detrimental influence of dissipation would be beneficial on the application level as well as on the fundamental level.

In spintronics---the spin version of electronics---the spin of electrons is usually manipulated as information carrier  \cite{AtsufumJMMM2020}; the spin of localized electrons in insulating magnets and the spin of delocalized electrons in metallic magnets. In both cases, the dissipation of the spins directly determines the speed of magnetization switching \cite{LiuScience2012,FukamiNN2016} and the speed of magnetic texture propagation \cite{BeachNM2005,ParkinScience2008} and, in turn, it influences the efficiency of spintronic devices that code information either in the ferromagnetic states or in magnetic solitons. The dissipation also determines the lifetime of magnon excitations above the magnetic textures. So, the smaller the dissipation, the longer information will survive when coded in magnons and the lower energy waste will be in magnon transport \cite{ChumakNP2015}. The current knowledge of spin dissipation is mostly based on micromagnetics (macrospin-mesh) approximations, where all the spins inside a magnet interact with the environment independently \cite{Landau1935,Gilbert2004,Kampfrath2011,GomonayLTP2014,Selzer2016,Baltz2018}. In two-sublattice antiferromagnets, this independent coupling to the environment leads to simple (intra-)sublattice Gilbert damping. However, a phenomenological approach showed that one should also expect inter-sublattice Gilbert damping \cite{LiuPRM2017,yuan_proper_2019,AkashPRB2018}. As shown below, the inter-sublattice Gilbert damping arises from the coupling to a bath shared between the sublattice magnetizations.

\begin{figure}[b]
\begin{center}
\includegraphics[width=0.45\textwidth]{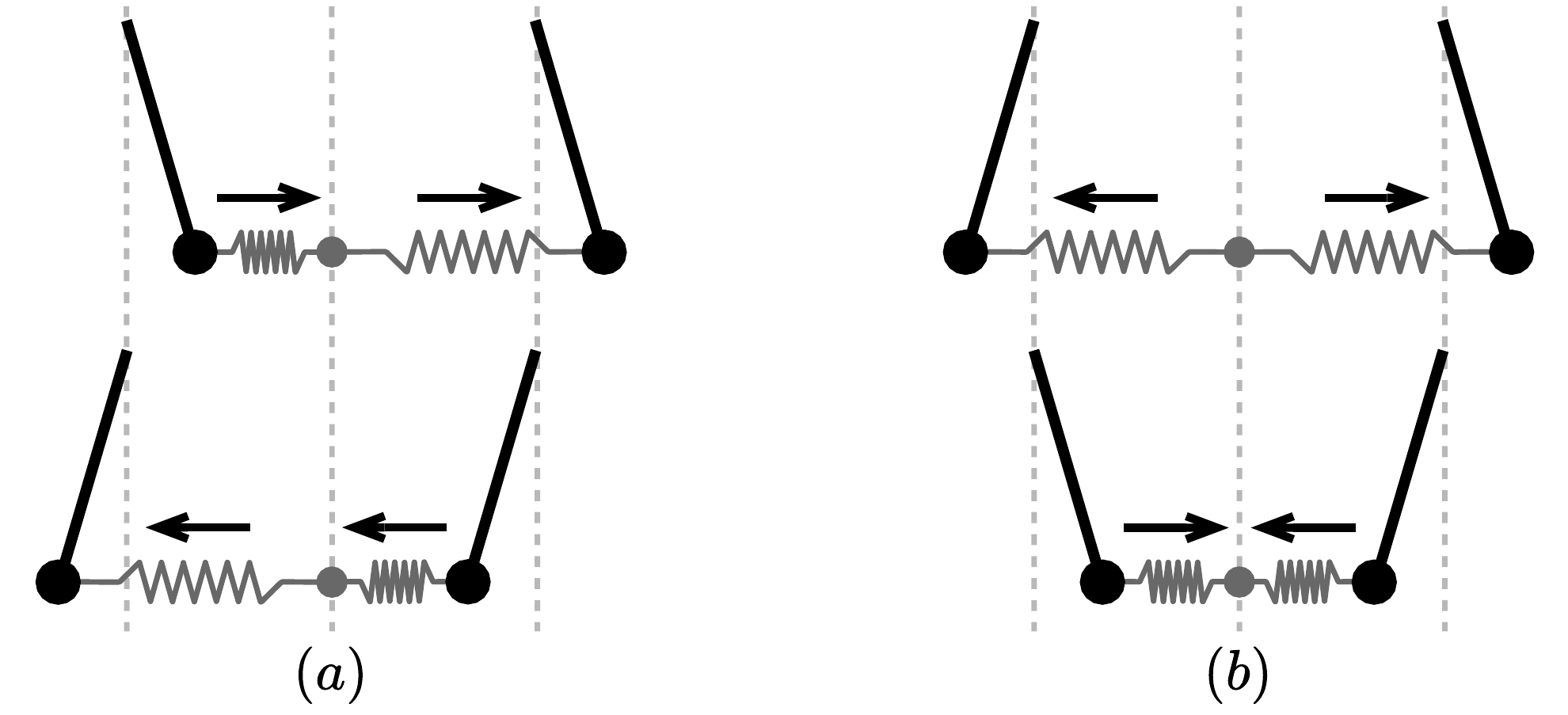}
\end{center}
\caption{Illustration of dissipative and dissipation-free modes for two coupled oscillators (illustrated as pendulums). The bath degrees of freedom are illustrated by the gray ball and the linear oscillator-to-bath coupling is illustrated by springs. The oscillation is between upper and lower figures. Oscillations of the center-of-mass coordinate (a) exert a force (black arrows) onto the bath degrees of freedom and, in turn, dissipates energy by exciting bath modes. In contrast, in relative-coordinate oscillations (b) the forces on the bath modes balance each other and, in turn, the bath modes are not excited and no energy is dissipated.} \label{fig: oscillators}
\end{figure}

\begin{figure}[b]
\begin{center}
\includegraphics[width=0.35\textwidth]{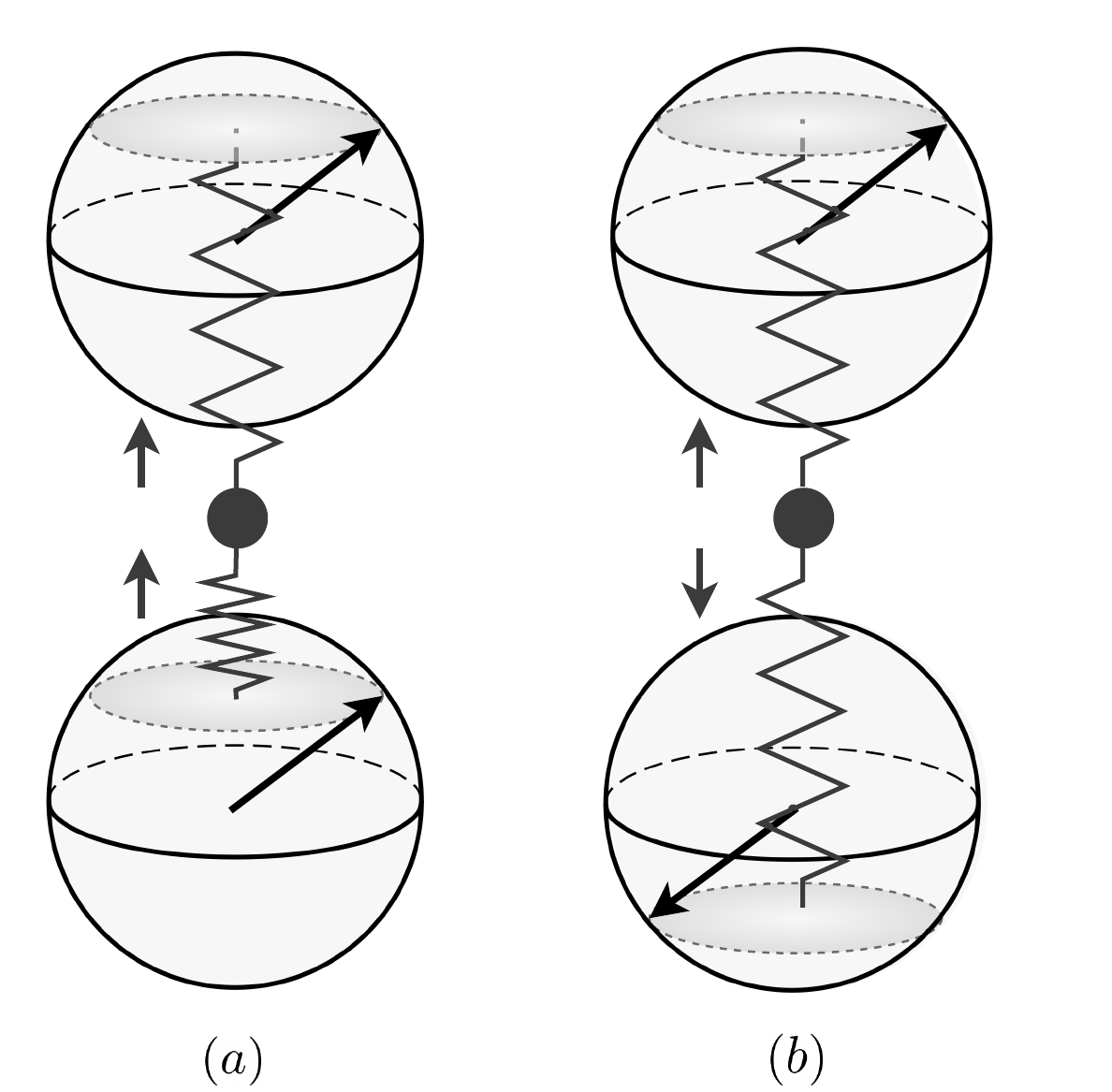}
\end{center}
\caption{Illustration of dissipative and dissipation-free modes for two coupled spins; for $z$-components only. The bath degrees of freedom are illustrated by the grey ball and the linear spin-to-bath coupling is illustrated by springs. The motion of the average magnetization (a) exerts a force (grey arrows) onto the bath degrees of freedom and, in turn, dissipates energy by exciting bath modes. In contrast, the motion of the Néel vector (b) does not excite the bath, as forces exerted onto the bath degrees of freedom always balance each other; in turn, the Néel vector is a dissipation-free mode.} \label{fig: spins}
\end{figure}

In this Letter, we study the dissipation of composite systems, where the subsystems share a common bath (reservoir). Using the Caldeira-Leggett approach \cite{caldeira_quantum_1983}, we integrate out the bath degree of freedoms. While each subsystem becomes dissipative, we identify a dissipation-free mode of the composite system; for illustration of the basic idea, see figures \ref{fig: oscillators} and \ref{fig: spins}. 
In particular, we consider a system of two coupled oscillators and a system of two coupled macrospins. In turn, we argue that dissipation-free (or low-dissipation) modes should appear in a large class of physical systems. Based on the dissipation-free modes, we explain how our results could be used to engineer low-dissipation information channels.

\textit{Two coupled oscillators.---}At first, we focus on the simpler system of two oscillators (not necessarily harmonic), which are coupled to each other and to a shared bath; see figure \ref{fig: oscillators}. The composite system of oscillators and bath is described by the Hamiltonian $\hat{H} = \hat{H}_o + \hat{H}_c + \hat{H}_{b}$, where $\hat{H}_o$ is the oscillators' Hamiltonian, $\hat{H}_b$ is the bath's Hamiltonian, and $\hat{H}_c$ is the Hamiltonian describing the oscillator-to-bath coupling. Here the oscillators' Hamiltonian is given by the sum of kinetic and potential energies, $\hat{H}_{o} =  \hat{p}_1^2/2m + \hat{p}_2^2/2m + V(\hat{q}_1, \hat{q}_2)$, where $\hat{q}_1, \hat{p}_1$ and $\hat{q}_2, \hat{p}_2$ are the position and momentum operators of oscillator $1$ and $2$ respectively and $V(\hat{q}_1, \hat{q}_2)$ is the oscillators' potential energy (including their interaction). For simplicity, we assumed the oscillators to have equal mass $m$.
Following the Caldeira-Leggett approach \cite{caldeira_quantum_1983}, we assume a linear oscillator-to-bath coupling and model the bath by harmonic oscillators. Explicitly, $\hat{H}_{b} = \sum_i ( \hat{p}^2_i/2m_i + m_i \omega_i^2 \hat{x}^2_i/2)$, where $\hat x_i, \hat p_i$ are position and momentum operators of the $i$-th bath oscillator  \footnote{The momenta of the harmonic bath-oscillators $\hat p_i$ must not be confused with the momenta of the oscillators $\hat p_1$ and $\hat p_2$. Such a confusion should never arise, as we don't need to write out the sum over $i$. If one would want to write the sum over $i$ explicitly, one could let it start from $i=3$ to avoid confusion.}; correspondingly, $m_i$ and $\omega_i$ are mass and frequency of the $i$-th bath oscillator. The coupling between the two oscillators and the bath modes is described by $\hat{H}_c = \sum_i \gamma_i\,  \hat{x}_i \hat{q}_1 +\sum_i \gamma_i\,  \hat{x}_i \hat{q}_2$; for simplicity, we assumed the coupling coefficients $\gamma_i$ to be the same for both oscillators \footnote{Here, we assume equal mass and equal coupling constants for both oscillators, such that the modes that decouple the kinetic and potential part of the action correspond to the dissipative and dissipation-free mode. However, the general requirement is that the ratio of the particles' masses and the ratio of coupling constants are equal: $\frac{m_2}{m_1}=\frac{\gamma_2}{\gamma_1}$. This may be used in experimental set-ups to realize a (nearly) dissipation-free mode for non-identical particles.}.

Using the Keldysh formalism in its path-integral version \cite{kamenev_field_2011, altland_condensed_2010} and integrating out the (Gaussian) bath degrees of freedom, we obtain the Keldysh partition function $Z = \int Dq_1 Dq_2\,  \mathrm{exp}[i\mathcal{S}]$ with the action
\begin{equation}\label{eq:action_oscillators}
\begin{split}
    &\mathcal{S} = \oint_K \mathrm{d}t \left(\frac{m}{2}(\dot{q}_1^2 + \dot{q}_2^2) - V(q_1,q_2)\right) \\
    &+ \oint_K\! \mathrm{d}t \oint_K\! \mathrm{d}t'\, \big(q_1(t) + q_2(t)\big) \alpha(t-t') \big(q_1(t') + q_2(t')\big)\, .
\end{split}
\end{equation}
The kernel function $\alpha(t-t')$ contains all the information about the coupling to the bath; the information about dissipation is contained in the retarded and advanced parts $\alpha^{R/A}(t-t')$, while the information about fluctuations is included in the Keldysh part $\alpha^K(t-t')$. Focusing on the noiseless dynamics, we can disregard the Keldysh part \footnote{The retarded and advanced propagator carry all information regarding the dissipative dynamics, as long as the system does not possess multiplicative noise, which can lead to a drift term in the deterministic part of the dynamics; see \cite{kamenev_field_2011}.}. For the retarded and advanced parts, we find $\alpha^{R/A}(\omega) = \int_{-\infty}^{ \infty } \frac{\mathrm{d}\epsilon}{2\pi}\ \epsilon J(\epsilon)/[\epsilon^2 - (\omega \pm i\eta)^2]$ with infinitesimal $\eta$ and the bath spectral density $J(\epsilon) = \pi \sum_i (\gamma_i^2/\omega_i)\, \delta(\epsilon - \omega_i)$; compare \cite{kamenev_field_2011}.

The form of the action, equation \eqref{eq:action_oscillators}, suggests to introduce the center-of-mass coordinate $Q = (q_1 + q_2) /2$ and the relative coordinate $q = q_1 - q_2$; then, only the center-of-mass coordinate appears in the second line of equation \eqref{eq:action_oscillators}. Thus, only the center-of-mass coordinate is affected by the coupling to the bath. In turn, the relative coordinate will be free from dissipation induced by the coupling to the bath; that is, it is a dissipation-free mode. 
Does that mean the relative coordinate, when excited, can never relax? The answer to this question is more subtle. To answer it, we consider the quasi-classical dynamics of center-of-mass and relative coordinates.

The quasi-classical dynamics of the center-of-mass coordinate $Q$ and the relative coordinate $q$ are found as follows: we start from the action, equation \eqref{eq:action_oscillators}; then, we rewrite the coordinates as $q_1 = Q + q/2$ and $q_2 = Q - q/2$; finally, we demand that the variations of the action with respect to quantum components of $Q$ and $q$ both vanish \cite{kamenev_field_2011}. Taking the variation with respect to quantum components in Keldysh formalism corresponds to varying the action in classical Lagrangian mechanics. In turn, we obtain the quasi-classical equations of motion,
\begin{subequations}
\begin{align}
     m \ddot{Q} &= - \frac{1}{2} \frac{\partial}{\partial Q}\bar V(Q, q) - \alpha_0 \dot{Q}\ , \label{eq:eom_Q} \\
     m \ddot{q} &= - 2\frac{\partial}{\partial q} \bar V(Q, q)\ ; \label{eq:eom_q}
\end{align}
\end{subequations}
for compact notation, we introduced $\bar V (Q, q) = V(Q + q/2, Q - q/2)$ and, for simplicity, we assumed the bath to be ohmic $J(\epsilon) \approx \alpha_0\, \epsilon\, \Theta(\epsilon)$, where $\Theta(\epsilon)$ is the Heaviside-theta function and $\alpha_0$ is some damping constant.

For the center-of-mass coordinate, the quasi-classical equation of motion \eqref{eq:eom_Q} contains a friction term  $\alpha_0 \dot Q$; thus, $Q$ is a dissipative coordinate. In contrast, for the relative coordinate, the quasi-classical equation of motion \eqref{eq:eom_q} contains no friction term; in turn, we call $q$ a dissipation-free coordinate. Note, however, that via the potential $\bar V (Q, q)$ the dissipation-free relative coordinate $q$ can be coupled to the dissipative center-of-mass coordinate $Q$. So, energy stored in the relative coordinate $q$ can be dissipated into the bath indirectly by exciting the center-of-mass coordinate $Q$ which, in turn, dissipates some of its energy into the bath. 

In the important special case of two identical linearly-coupled harmonic oscillators $V(q_1, q_2) =  m \omega_0^2 q_1^2 + \gamma q_1 q_2 + m \omega_0^2 q_2^2$, the center-of-mass and relative coordinates decouple $\bar V(Q,q) = (2 m \omega_0^2 + \gamma) Q^2 + (m \omega_0^2/2 - \gamma/4) q^2$. In turn, the coordinates $Q$ and $q$ describe the eigenmodes of the systems and the relative coordinate becomes a truly non-dissipative mode. This result of having a truly non-dissipative mode is not restricted to the case of identical linearly-coupled harmonic oscillators but holds in more general case, when center-of-mass motion and relative motion decouple; that is, for $\bar V(Q,q) = V_Q(Q) + V_q(q)$ with arbitrary potentials $V_Q(Q)$ and $V_q(q)$. While this case might seem artificial at first, a similar situation naturally arises in two-sublattice antiferromagnets as we shall see below.

\textit{Two-sublattice antiferromagnet.---}Anti\-ferro\-mag\-nets are conventionally modeled as a composite system of ferromagnetic sublattices \cite{KittelPR1951,Keffer1952,Kampfrath2011,GomonayLTP2014,Selzer2016,Baltz2018}. Here, we focus on a two-sublattice antiferromagnet, where the sublattice magnetizations are coupled to a shared bath.

The composite system of sublattice magnetizations and bath is described by the Hamiltonian $\hat{H} = \hat{H}_{s}+ \hat{H}_c + \hat{H}_{b}$, where $\hat{H}_{s}$ is the Hamiltonian for the sublattice magnetizations, $\hat{H}_b$ is the bath Hamiltonian, and $\hat{H}_c$ is the Hamiltonian describing the coupling between the sublattice magnetizations and the bath.
We model the sublattice magnetizations by two large spins (macrospins) with corresponding spin operators $\mathbf{\hat S}_1 = ( \hat S_1^x, \hat S_1^y, \hat S_1^z)$ and $\mathbf{\hat S}_2 = ( \hat S_2^x, \hat S_2^y, \hat S_2^z)$. The coupling between the two sublattice magnetizations is then described by the exchange interaction $J\, \mathbf{\hat{S}_1 \cdot \hat{S}_2}$, where the exchange constant $J$ is positive to favor the antiferromagnetic order (anti-aligned macrospins). Also including a Zeeman-energy term to describe the coupling to an external magnetic field $\mathbf H$, the Hamiltonian for the sublattice magnetizations becomes $\hat H_\mathbf{S} =  J\, \mathbf{\hat{S}_1 \cdot \hat{S}_2} - \mathbf{H} \cdot (\mathbf{\hat{S}_1 + \hat{S}_2})$.
Following the Caldeira-Leggett approach \cite{caldeira_quantum_1983} again, we model the bath by harmonic oscillators and assume a linear coupling between the macrospins and the bath degrees of freedom.
Explicitly, $\hat{H}_{b} = \sum_i ( \mathbf{\hat{p}}^2_i/2m_i + m_i \omega_i^2 \mathbf{\hat{x}}^2_i/2)$, where $\mathbf{\hat x}_i, \mathbf{\hat p}_i$ are position and momentum operators of the $i$-th bath oscillator; correspondingly, $m_i$ and $\omega_i$ are mass and frequency of the $i$-th bath oscillator. The coupling between the two macrospins and the bath modes is described by $\hat{H}_c = \sum_i \gamma_i\,  \mathbf{\hat{x}}_i \cdot \mathbf{\hat{S}}_1 + \sum_i \gamma_i\,  \mathbf{\hat{x}}_i \cdot \mathbf{\hat{S}}_2$, where we assumed the coupling coefficients $\gamma_i$ to be the same for both macrospins.

Again, using Keldysh formalism \cite{kamenev_field_2011, altland_condensed_2010}, we integrate out the bath and obtain the Keldysh partition function $Z = \int Dg_1 Dg_2\,  \mathrm{exp}[i\mathcal{S}]$ with the action
\begin{align}
      \mathcal{S}& = -\oint_K\!\! \mathrm{d}t\, \big(i \bra{\dot{g}_1} \ket{g_1}\! +\!i \bra{\dot{g}_2}\ket{g_2}\! +\! J\mathbf{S}_1\! \cdot\! \mathbf{S}_2\! -\! \mathbf{H}\! \cdot\! (\mathbf{S}_1\!+\!\mathbf{S}_2) \big)\nonumber \\
 &+ \oint_K\!\! \mathrm{d}t \oint_K\!\! \mathrm{d}t'  \big (  \mathbf{S}_1(t)  +  \mathbf{S}_2(t)\big)  \alpha(t-t')\big (  \mathbf{S}_1(t')  +  \mathbf{S}_2(t')\big)\ ,
    \label{eq:action_spins}
\end{align}
where the spin vectors are given by $\mathbf{S}_1(t) = \langle g_1| \mathbf{\hat S}_1|g_1\rangle$ and $\mathbf{S}_2(t) = \langle g_2| \mathbf{\hat S}_2|g_2\rangle$ with the spin-coherent states $|g_1\rangle$ and $|g_2\rangle$; compare \cite{altland_condensed_2010}.
As in the previous section, the kernel function $\alpha(t-t')$ contains all the information about the coupling to the bath; the expression of the dissipative parts $\alpha^{R/A}(t-t')$ is the same as above and again we disregard the fluctuation part $\alpha^K(t-t')$.

As before, the structure of the action, equation \eqref{eq:action_spins}, suggests to introduce different coordinates irrespective of the precise form of $\alpha^{R/A}(\omega)$. Explicitly, the dissipative part of the action (second line of equation \eqref{eq:action_spins}) suggests to introduce the average (per-sublattice) magnetization $\mathbf M = (\mathbf{S}_1 + \mathbf{S}_2)/2$ and the Néel vector $\mathbf N = \mathbf{S}_1 - \mathbf{S}_2$ ; then, only the magnetization appears in the second line of equation \eqref{eq:action_spins}. So, intuitively, one can already expect from the action that only the magnetization will experience dissipation, while the Néel vector will be free from dissipation; see figure \ref{fig: spins}. This time, however, the situation is more complicated than for the two coupled oscillators. The reason is that the Berry-phase terms $i \bra{\dot{g}_1} \ket{g_1}$ and $i \bra{\dot{g}_2}\ket{g_2}$, which take the role of the kinetic energy for the two macrospins \cite{berry_quantal_1984}, cannot easily be written in terms of the spin vectors $\mathbf{S}_1$ and $\mathbf{S}_2$ \footnote{The Berry-phase terms can in fact be written as the coupling to an effective magnetic charge, see \cite{altland_condensed_2010}. However, for our calculations this is neither necessary nor convenient.}; in turn, the Berry-phase terms cannot be expressed in terms of $\mathbf{M}$ and $\mathbf{N}$. Nevertheless, as we will show, the intuitive expectation still holds: the magnetization is dissipative; the Néel vector is a dissipation-free mode.

To determine the Berry-phase terms $i \bra{\dot{g}_1} \ket{g_1}$ and $i \bra{\dot{g}_2}\ket{g_2}$, it is easiest to use the Euler-angle representation of spin-coherent states \footnote{We use the Euler-angle representation for spin-coherent states as described in \cite{altland_condensed_2010}. For $\ket{g_1}$, the Euler-angle representation is $\ket{g_1} = \mathrm{exp} [-i \phi_1 \hat S_1^z] \mathrm{exp} [-i \theta_1 \hat S_1^y] \mathrm{exp} [-i \psi_1 \hat S_1^z] \ket{\uparrow_1}$, where $\ket{\uparrow_1}$ is the eigenstate to $\hat S_z^1$ with maximal eigenvalue $S_1$; that is, $\hat S_z^1 \ket{\uparrow_1} = S_1 \ket{\uparrow_1}$. The coordinates $\phi_1, \theta_1$ are spherical-coordinate angles describing the spin orientation. The other angle $\psi_1$ represents a gauge freedom, which we fix as in \cite{shnirman_geometric_2015}; roughly speaking as $\dot \psi_1 = - \dot \phi_1$. The representation is analog for spin-coherent state $\ket{g_2}$ with indices $1$ replaced by $2$.}. This representation is quite intuitive, as it describes the spins as vectors in spherical coordinates $\mathbf S_1 = S_1\, (\sin \theta_1 \cos \phi_1, \sin \theta_1 \sin \phi_1, \cos \theta_1)$ and $\mathbf S_2 = S_2\, (\sin \theta_2 \cos \phi_2, \sin \theta_2 \sin \phi_2, \cos \theta_2)$ with spin lengths $S_1$ and $S_2$ and spherical angles $\theta_1, \phi_1$ and $\theta_2, \phi_2$. For the Berry-phase terms, we find $i \bra{\dot{g}_1} \ket{g_1}= S_1\, (1 - \cos \theta_1) \dot \phi_1$ and $i \bra{\dot{g}_2} \ket{g_2}= S_2\,  (1 - \cos \theta_2) \dot \phi_2$. We can then derive quasi-classical equations of motion by varying the action \footnote{Explicitly, we vary the action with respect to the quantum components of the coordinates and demand that the variation vanishes.}.

After varying the action, we obtain for the quasi-classical equations of motion $ \mathbf{\dot{S}}_1 = \mathbf{S_1}\times (\mathbf{H}-J\mathbf{S_2} - \alpha_0 \mathbf{\dot{S}}_1 -\alpha_0\mathbf{\dot{S}}_2) $ and, analogously, $\mathbf{\dot{S}}_2 = \mathbf{S_2}\times (\mathbf{H}- J\mathbf{S_1}  - \alpha_0 \mathbf{\dot{S}}_2 -\alpha_0\mathbf{\dot{S}}_1)$, where for simplicity we assumed the bath to be ohmic $J(\epsilon) \approx \alpha_0\, \epsilon\, \Theta(\epsilon)$ again. These two equations are Landau-Lifshitz-Gilbert equations for each sublattice.
In both equations, the first term describes the precession around the external field, the second term describes the precession around the other spin due to the exchange interaction, the third term describes the usual (intra-sublattice) Gilbert damping \cite{gilbert_phenomenological_2004}, and the fourth term describes the inter-sublattice Gilbert damping that was previously described phenomenologically \cite{yuan_proper_2019}. We can now combine the sublattice-Landau-Lifshitz-Gilbert equations into equations of motion for the average magnetization and the Néel vector \footnote{Note that the magnitude of $\mathbf{M}$ does not relax according to \eqref{eq: magnetization}, which is due to the permutation symmetry of the two sublattices in an antiferromagnet. An inhomogenous field acting on the two sublattices \cite{ZeleznPRL2014} or coupling between the sublattice magnetization and the non-shared bath \cite{yuan_proper_2019} can break the symmetry and thus induce the relaxation of $\mathbf{M}$ toward zero.}
\begin{subequations}
\begin{align}
    \mathbf{\dot{M}} &= \mathbf{M}\times (\mathbf{H}-2\alpha_0   \mathbf{\dot{M}})\ , \label{eq: magnetization}
    \\
    \mathbf{\dot{N}} &=  \mathbf{N}\times( \mathbf{H}-2\alpha_0 \mathbf{\dot{M}}-2J \mathbf{M})\ . \label{eq: neel vector}
\end{align}
\end{subequations}
For the average magnetization $\mathbf{M}$, the quasi-classical equation of motion \eqref{eq: magnetization} contains a Gilbert-damping term $-2\alpha_0  \mathbf{M}\times \mathbf{\dot{M}}$; thus, it is a dissipative coordinate. In contrast, Néel vector is a dissipation-free coordinate. Although the term $-2\alpha_0  \mathbf{N}\times \mathbf{\dot{M}}$ is clearly induced by the coupling to the bath, it is not dissipative, as can be understood from the dynamics. As described by equation \eqref{eq: magnetization}, the magnetization precess around the magnetic field but, due to the Gilbert-damping term, it will relax towards it. After some time, the magnetization will be aligned to the magnetic field and will not move anymore, $\mathbf{\dot M} = 0$. Then, as described by equation \eqref{eq: neel vector}, the Néel vector continues to precess around $\mathbf H - 2 J \mathbf M$; it will not stop precessing, as it cannot dissipate into the shared bath.


\textit{Comparison and generalizations.---}In both cases, coupled oscillators and coupled macrospins, we found one dissipative and one dissipation-free mode. The dissipation-free modes are those modes that do not excite the bath because of the forces exerted on the bath modes balance each other; see figures \ref{fig: oscillators} and \ref{fig: spins}. The dissipation-free modes we found are: the relative coordinate for two coupled oscillators; and the Néel vector for two coupled spins.

The existence of dissipation-free modes can be traced back to the linear coupling between the subsystems and the shared bath. Explicitly, we can rewrite $\hat{H}_c = \sum_i \gamma_i\,  \hat{x}_i \hat{q}_1 +\sum_i \gamma_i\,  \hat{x}_i \hat{q}_2 =  \sum_i \gamma_i\,  \hat{x}_i (\hat{q}_1 + \hat{q}_2)$, which makes it clear that only the center-of-mass coordinate $Q$ couples to the shared bath but not the relative coordinate $q$. Similarly, for the antiferromagnet, we can rewrite $\hat{H}_c = \sum_i \gamma_i\,  \mathbf{\hat{x}}_i \cdot \mathbf{\hat{S}}_1 + \sum_i \gamma_i\,  \mathbf{\hat{x}}_i \cdot \mathbf{\hat{S}}_2 = \sum_i \gamma_i\,  \mathbf{\hat{x}}_i \cdot ( \mathbf{\hat{S}}_1 + \mathbf{\hat{S}}_2)$, which makes it clear that only the magnetization $\mathbf M$ but not the Néel vector $\mathbf N$ couples to the shared bath. Note that, the dissipation-free modes are not affected by the bath type (ohmic or non-ohmic); it only affects the dissipation of the other mode. So, we can expect to find dissipation-free modes in a large class of systems; namely, whenever identical subsystems couple to only a single bath that is shared between all them \footnote{While we have focused on systems composed of two subsystems, the generalization to $N$ identical subsystems with a single shared bath is straightforward. Interestingly, one finds only $1$ dissipative mode and $N-1$ dissipation-free modes.}.

While the described class of systems with dissipation-free modes is already large, the existence of dissipation-free modes is not restricted to that class. As one example, consider two coupled spins, where $|\mathbf{S}_1| = 2 |\mathbf{S}_2|$ but the spin-bath coupling $\gamma_i^{(1)} = \gamma_i^{(2)}/2$; then, $\mathbf S_1 + \mathbf S_2$ is dissipative, whereas $\mathbf S_1 - \mathbf S_2$ is dissipation free. For spintronics this example is particularly interesting, because it might explain why certain ferrimagnets have a low dissipation.

\textit{Application.---}To understand how our results can be applied, note at first that baths are not necessarily shared between subsystems of a composite system. For non-shared baths, the subsystems cannot compensate each other's "force" onto the bath modes and, in turn, one cannot expect a dissipation-free mode. So, a mode that would be dissipation-free for a shared bath becomes dissipative in real systems in two ways: indirectly insofar it couples to the dissipative mode; or directly insofar the subsystems couple to separate baths. As a result, in a composite system we can reduce the dissipation of some modes also in two ways: either by engineering the system to manipulate the coupling between its eigenmodes; or by engineering the baths (with the subsystem-to-bath couplings) such that the subsystems mostly share the same bath.

Our results even suggest a way to engineer a low-dissipation mode in composite systems: first, the baths and the subsystem-to-bath couplings must be engineered in such a way that the subsystems mostly share one bath; secondly, the subsystem-to-subsystem coupling must be engineered such that one of the systems' eigenmodes agrees with the dissipation-free mode we expect from the system-to-bath coupling. While this might seem artificial at first, it might naturally occur in two-sublattice antiferromagnets: first, the sublattices are very close to each other, which makes it likely that they mostly couple to a shared bath; second, in the exchange interaction between the sublattice spins the average magnetization and the Néel vector decouple from each other. In turn, even in real antiferromagnets, the Néel vector is an almost non-dissipative mode. So, our simple model offers an explanation for why, in some antiferromagnets, the dissipation of the Néel vector is so much smaller than the dissipation of the magnetization.
In turn, our results also offer an explanation for the large magnon lifetimes typically found in some antiferromagnets and layered magnets \cite{Hermanny2021,LiuPRM2017}.

\textit{Conclusion.---}For two model systems (coupled oscillators and coupled macrospins), we derived the dissipative quasi-classical equation of motion, while assuming that the subsystems are coupled to a shared bath. In both cases, we identified a dissipation-free mode (relative coordinate and Néel vector). Generalizing those results, we argued that a dissipation-free mode should be expected in a broad class of model systems; namely, whenever subsystems of a composite system are coupled to a shared bath with linear subsystem-to-bath coupling. When the dissipation-free mode agrees with an eigenmode of the system, then there is not even an indirect dissipation channel and one can expect a very long lifetime for an excitation of that mode.

Finally, we discussed how our results could be used in real systems to engineer low-dissipation modes. Such low-dissipation modes could be exploited to design an efficient information transfer channel through a system, as the corresponding excitations will have a very long lifetime. Similarly, exploiting a low-dissipation mode could be used to reduce decoherence; for a known special case for qubits, see \cite{Zanardi1997,duan_reducing_1998,Kielpinski2001}. Two-sublattice antiferromagnets seem to be candidates for systems in which a low-dissipation mode can naturally occur. In turn, our results also offer an explanation for the low Néel-vector dissipation and, correspondingly, for the large magnon lifetime in some antiferromagnets \cite{LiuPRM2017}.

\textbf{\textit{Acknowledgments.---}}
We thank A. Shnirman for a fruitful discussion. H.Y.Y acknowledges the European Union's Horizon 2020 research and innovation programme under Marie Sk{\l}odowska-Curie Grant Agreement SPINCAT No. 101018193. R.A.D. is member of the D-ITP consortium that is funded by the Dutch Ministry of Education, Culture and Science (OCW). R.A.D. has received funding from the European Research Council (ERC) under the European Union's Horizon 2020 research and innovation programme (Grant No. 725509). This work is part of the research programme Fluid Spintronics with project number 182.069, which is (partly) financed by the Dutch Research Council (NWO).

\bibliography{references.bib}

\end{document}